\documentclass[runningheads]{svmult}

\usepackage{makeidx}   
\usepackage{graphicx}  
\usepackage{subeqnar}  
\usepackage{multicol}  
\usepackage{physprbb}  
\makeindex             

\begin{document}
\title*{Properties of Spiral and Elliptical Galaxy 
Progenitors at $z > 1$}
\toctitle{Proto-Disk and Elliptical Properties}
%
%
\titlerunning{Proto-Disk and Elliptical Properties}
%
\author{Christopher J. Conselice}
\authorrunning{Christopher J. Conselice}
%
%
\institute{California Institute of Technology, Pasadena, CA, USA}

\maketitle              

\begin{abstract}

We present the results of a Hubble Space
Telescope and ground-based optical and near-infrared 
study to identify progenitors of spirals and ellipticals at $z > 1$.  
We identify these systems through photometric and spectroscopic redshifts,
deep K-band imaging, stellar mass measurements, and high resolution
imaging.  The major modes of galaxy formation, including  major mergers, minor
mergers, and accretion of intergalactic gas, and their relative 
contributions towards building up the stellar masses of galaxies, can now 
be directly measured using these data.

\end{abstract}

\section{Introduction}

The generally accepted modern hierarchical galaxy formation picture consists
of galaxies forming in dark matter halos that later merge to form larger
halos and more massive galaxies.  The end result of this evolution is
the morphological and stellar population mix in the nearby
universe.  This picture, however, remains largely untested.  Understanding 
how the modern galaxy population was put into place 
requires understanding when and how stars
(and hence galaxies) formed.  Based on several decades of observations
and modeling of stellar evolution we know that the stars in nearby
galaxies contain a wide diversity of ages and metallicities. To
first order these differences correlate with galaxy type and
environment. Generally, early-types or elliptical galaxies
are dominated by old stars and are found in dense environments, while later 
type galaxies have a mix of young and old stellar populations and are found
in lower density areas.

\begin{figure}[b]
\begin{center}
\hspace{-1.6cm}
\includegraphics[width=1.1\textwidth]{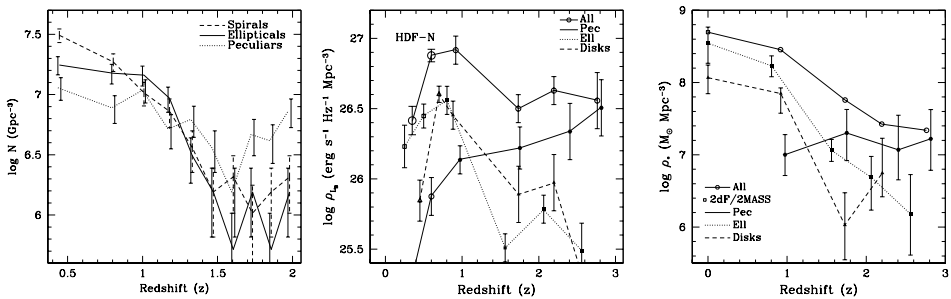}
\end{center}
\caption[]{The relative co-moving number, rest-frame B-band luminosity, and
stellar mass density of galaxies as a function of redshift from deep 
NICMOS images
of the Hubble Deep Field North (Conselice et al. 2004a).  Points
at redshifts $z < 0.5$ are taken from Brinchmann \& Ellis (2000), 
Fukugita et al. (1998) and the 2dF/2MASS survey.}
\label{eps1}
\end{figure}

Directly measuring the galaxy mass assembly and star formation history has
now been accomplished out to $z \sim 3-6$.  However, these measurements
do not tell us {\em how} galaxies formed.  One way to address
this question is to include high resolution imaging, such as from deep Hubble 
Space 
Telescopes (HST) images.  Imaging surveys with HST show that 
galaxies evolve
into normal systems from peculiars between $z \sim 1-2$ ($\sim 10$ Gyrs
ago) (Figure~1; e.g., van den Bergh et al. 2001; Conselice et al. 2003). 
At redshifts 
higher than $z \sim 1.5$ most galaxies are distorted
and asymmetric (Abraham et al. 1996; Conselice
et al. 2003).  Deep NICMOS observations of the Hubble Deep Field North, 
which samples the rest frame B-band morphologies
of galaxies at $z > 1.2$, demonstrates that these galaxies are
intrinsically distorted in the rest-frame optical, and that we are not 
witnessing morphological k-correction effects (Papovich et al. 2003). 

These 
high redshift galaxies are also undergoing large amounts of star formation 
(e.g., Madau et al. 1998), creating the normal bright galaxies we see today.  
What causes the structural peculiarities in these galaxies, and presumably
also the induced star formation? If we can
answer this it will reveal the formation modes of galaxies, and allow
us to quantify the relative contributions of different formation 
processes.

\begin{figure}[b]
\begin{center}
\hspace{-0.8cm}
\includegraphics[width=0.7\textwidth]{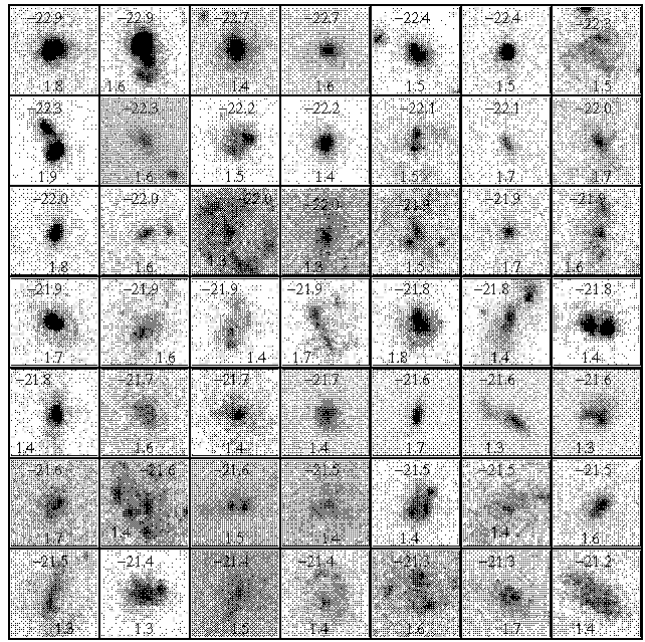}
\end{center}
\caption[]{\footnotesize Examples of galaxies in ACS GOODS images  
whose photometric redshifts place them at $1 < z < 2$.  These
are ordered from brightest to faintest down to M$_{\rm B}$ = $-21$.  The
upper number is the M$_{\rm B}$ of each galaxy and the lower number is its
redshift.  There is  a large
diversity of properties, from systems that appear very peculiar
to those that look similar to normal galaxies.   Scale of these images is 
$\sim 2''$  on each side, which 
corresponds to about 17 kpc at these redshifts.}
\vspace{-1in}
\end{figure}

\begin{figure}[b]
\begin{center}
\hspace{-1.4cm}
\includegraphics[width=0.8\textwidth]{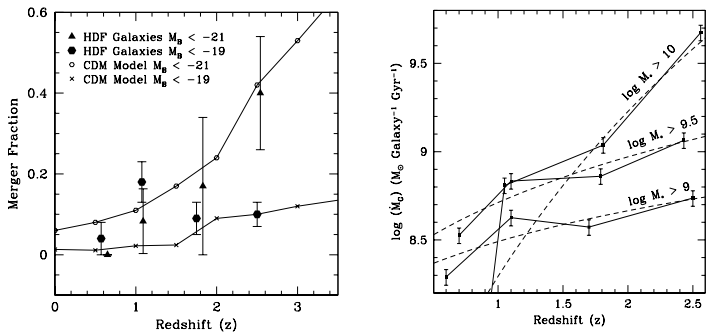}
\end{center}
\caption[]{Left panel: major merger fractions to $z \sim 3$ at
magnitude limits M$_{\rm B} = -21$ and $-19$. Semi-analytical model
predictions are also shown.  Right panel: Stellar mass accretion
history from major mergers as a function of initial mass 
(see Conselice et al. 2003). }
\label{eps1}
\end{figure}

\begin{figure}[b]
\begin{center}
\hspace{-1.6cm}
\includegraphics[width=0.8\textwidth]{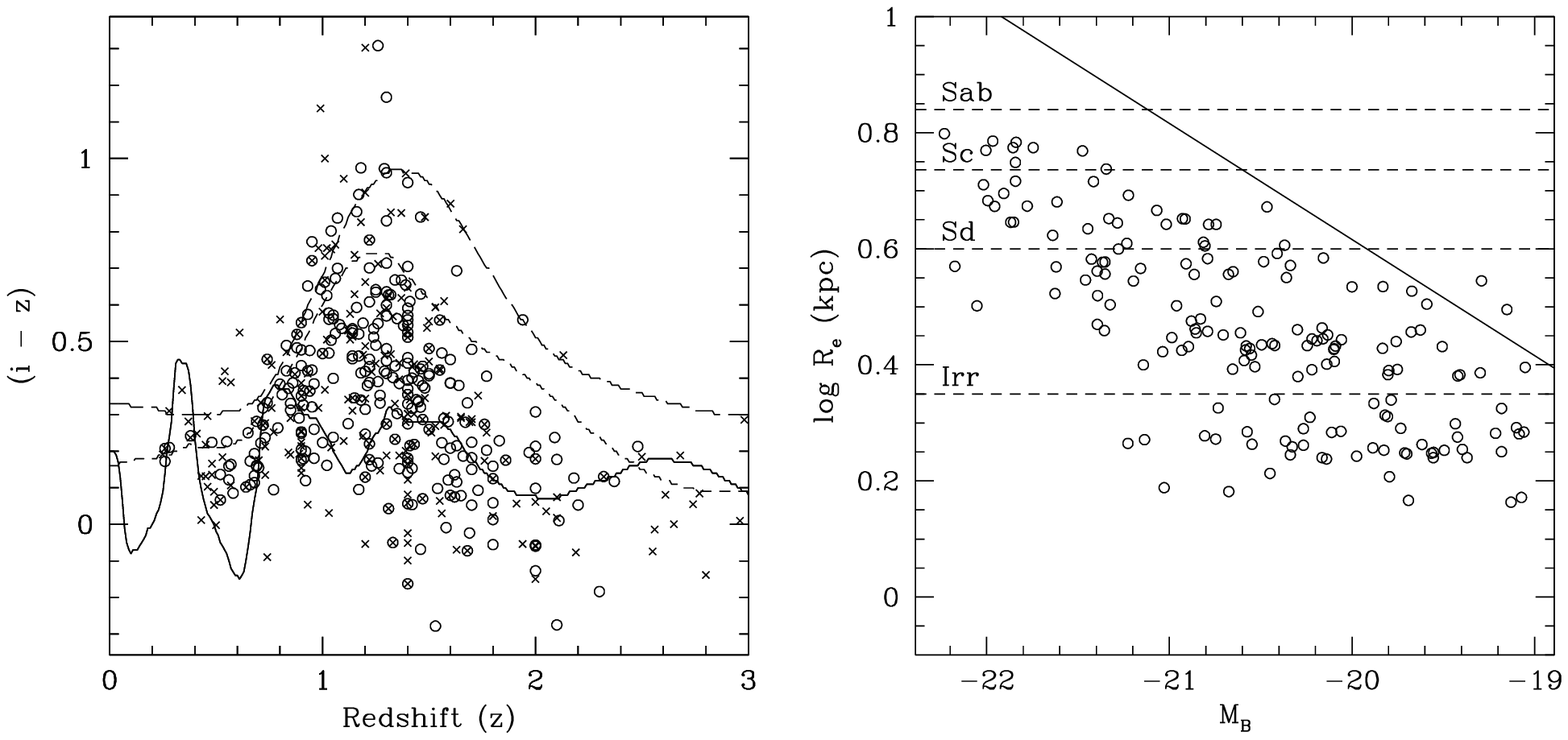}
\end{center}
\caption[]{Left panel: The distribution of $(i-z)$ colors for LDOs (circles) 
as a function of 
redshift with two Coleman, Wu and Weedman spectral energy distributions and a 
Kinney et al.
starburst model plotted (see Conselice et al. 2004b).  These are from bluest to
reddest - starburst (solid line), Scd (dashed), Sbc (long dashed).  Right
panel:  Absolute
 magnitude effective radius relationship for LDOs. The solid
line is the canonical Freeman disk relationship at $z \sim 0$.  The dashed
horizontal lines show the effective radii of different nearby galaxy types. }
\label{eps1}
\end{figure}

\section{Mass Assembly as a Function of Morphology}

The first step towards understanding the formation mechanisms of galaxies
is to have a robust determination of how the structures of galaxies
change with redshift, and how these changes relate
to the star formation and mass assembly history.  Figure~1 shows 
co-moving number, luminosity, and stellar mass densities of ellipticals, 
spirals, and peculiar galaxies as a function of
redshift.  One interesting trend from this figure, besides the dominance
of spirals and ellipticals at $z < 1.5$ and peculiars at $z > 1.5$, is
the fact that there exists an equilibrium point at
$z \sim 1.5$ where the relative fraction of luminosity, mass and
number densities for normal galaxies (disks/ellipticals) and peculiars
are nearly equal.  In all regards this is the redshift in
galaxy evolution where modes of forming galaxies are
rapidly transitioning.  This trend is also seen when studying
 the NICMOS observations of the Hubble Deep Field North and in the Hubble 
Deep Field South.  

There is a growth in both
the stellar mass density and number density of spirals and ellipticals
at $z < 1$, with a corresponding decrease in the number of peculiars
(see also Brinchmann \& Ellis 2000). Star formation is still occurring
in disks and ellipticals at $z < 1$ as the stellar mass densities for
these types grows by a factor of two or more (Figure~1).  The luminosity 
density for both ellipticals and spirals also
peaks at $z \sim 1$, and 
mass to light ratios for these systems increase with lower redshift at 
$z < 1$.  At $z > 2$ peculiar galaxies
consistent with major merging (Conselice et al. 2003) are dominating the
luminosity and stellar mass density, suggesting that this is the
dominate star formation process at early times.

\section{Tracing Galaxy Formation}

There are four major global methods by which galaxies can form, or how
star formation is triggered.  One is
through major mergers where galaxies are built up by merging with
systems of similar mass.  The other is through minor mergers where
the mass of a galaxy is built up by the accretion of less massive
galaxies.  The third method is through accretion of intergalactic
gas which forms stars around galaxies in disk like
structures (Abadi et al. 2003).  A fourth is through tidal interactions with 
nearby galaxies.
This is perhaps an over simplification of the star formation
triggering process, but it is a valuable starting point for understanding
galaxy formation. 

The quantitative structural properties of galaxies
reveal the formation mechanisms responsible for producing
galaxy and star formation (e.g., Conselice 2003).  We can use the structures 
of galaxies, including
their sizes and stellar masses, to argue which systems at high
redshifts are likely progenitors of either disks or spheroidal
components. Figure~2 shows the brightest galaxies at $1 < z < 2$ selected
in the near infrared in the GOODS south field (Giavalisco et al. 2004).
These systems must be the progenitors of disks and ellipticals and many show
structures suggestive of this.

\subsection{\bf Major Mergers - Spheroid Formation}

Major mergers occur when two galaxies of similar mass merge together to form 
a more massive system.
Galaxies which are undergoing major mergers can be identified in the
rest-frame optical through their large asymmetries in comparison
to other parameters, such as color or clumpiness (Conselice et al. 2000; 
Conselice 2003).  We can use the
asymmetry index to
determine how common major mergers are at various redshifts and to measure
properties of these mergers.    The results of this are shown
in Figure~3 which plots the inferred merger fraction as a function of
redshift for galaxies at two luminosity limits.  It appears
that most major mergers at high redshift occur in 
luminous systems.  This is also the case for the most massive
galaxies at the same redshifts (Conselice et al. 2003). About 50\% of the 
most luminous
and massive galaxies at high redshift are undergoing major mergers.  The merger
rate for the most luminous and massive systems however rapidly declines
at lower redshifts.   A smaller fraction of fainter and less luminous
systems are undergoing mergers, suggesting that some other process is
responsible for their formation.

 Do these mergers add enough material
to produce a massive galaxy at low redshift from the most massive and
brightest systems found at $z > 2$?  This is uncertain, as
it depends not only on the merger rate, but also on the induced star
formation produced through the merger.  Integrating the mass obtained
through major mergers (Figure~3) with reasonable and empirical assumptions 
about 
star formation histories induced by mergers, it appears that massive 
ellipticals can
be formed through multiple merger and starbursts.  
However, a large fraction of stars cannot
have formed through major mergers, given that the fraction of galaxies
involved in major mergers declines rapidly at $z < 2$ and stellar
mass is still assembled at these redshifts (e.g., Dickinson et al. 2003).
Figure~3 also shows that the measured merger fraction is lower at low-$z$ 
than what is predicted
in CDM models, such that there are too many major mergers produced
in these models at low redshift.  That is, massive galaxy formation occurs 
earlier than predicted in semi-analytic CDM models.

\subsection{\bf Minor Mergers}

Minor mergers are likely a major method for building up the masses
of galaxies, as predicted in simulations (Murali et
al. 2002).   The observed role
of minor mergers is however not know with certainty, simply
because it is difficult to identify them, as they do not produce
major structural effects on galaxies.  One clue that minor
mergers might be occurring in large numbers at $z < 1$ is that about
50\% of the total stellar mass in the universe appears to form at $z < 1$;
yet major mergers are not producing this increase (Conselice 
et al. 2003).   The properties of minor mergers can be studied out
to $z \sim 1$ most effectively using pair studies of galaxies
either in the optical (Patton et al. 2002) or in the near infrared
(Bundy et al. 2004).  The optical pair fraction suggests that minor
merger rates are high enough to produce suitable increases in stellar mass.
The
merger rate is, however, lower when studying pairs in the near infrared
(Bundy et al. 2004).  The reason for this is that pre-accreted satellite 
galaxies are star forming with low mass to light ratios, resulting in a lower
near infrared pair fraction.  
The amount of stellar mass added to galaxies through minor mergers at $z < 1$ 
is roughly equal to
the gain in stellar mass at these redshifts, although
this is likely a coincidence (Bundy et al. 2004).  However, if these 
merging/interacting satellites are creating new stars with a total stellar 
mass equal to that of the original 
satellite, minor mergers would produce enough mass through tidally
induced star formation to account for the observed growth in stellar
mass.  Detailed determinations of the star formation and stellar mass
build-up induced in
pairs of galaxies at $z < 1$ are necessary to quantify the importance of
this effect.

\subsection{\bf Disk Formation}

Another major type of galaxy formation is through smooth accretion of
intergalactic gas.  Morphologically,
systems undergoing accretion will appear as disks, with a relatively high 
amount of
star formation than what is found in modern spirals.  Some 
forming disks have possibly
been found in deep HST imaging in the Hubble Deep Field South
(Labb{\'e} et al. 2003) and the GOODS South field (Conselice et al. 2004b).
These are typically found between redshifts $1 < z < 2$ and are symmetric
galaxies containing bright outer regions with low central light
concentrations.  
These bright regions create
 unconcentrated  structures, which is how these
systems are identified. Figure~2 shows examples of these luminous diffuse 
objects 
(LDOs) found in the GOODS South field (see Conselice et al. 2004b).

These systems have star formation rates consistent with
starbursts (Figure~4) with typical uncorrected for extinction UV
derived star formation rates of $\sim 4$~M$_{0}$ yr$^{-1}$. The integrated 
star 
formation rate in these galaxies accounts for 35-40\% of the total
star formation rate between redshifts $1.5 < z < 2.5$.  These systems 
also have sizes and scaling relationships consistent
with disk galaxies in formation (Figure~4). Some systems also contain
structures such as spiral arms and bars (Labb{\'e} et al. 2003; Conselice
et al. 2004b).  The star formation rate peak found between $1 < z < 2$
appears to relate to the formation of these disk-like systems through
accretion, while at lower/higher redshifts, minor/major mergers are the
dominate processes.

%

\end{document}